\documentclass[12pt]{iopart}

\usepackage{graphicx}
\begin{document}

\title[Nonlinear optics in graphene]{Nonlinear optics of graphene in a strong magnetic field}

\author{Xianghan Yao, Alexey Belyanin}

\address{Department of Physics \& Astronomy, Texas A\&M University, USA}
\ead{belyanin@tamu.edu}
\begin{abstract}

Graphene placed in a magnetic field possesses an extremely high mid/far-infrared optical nonlinearity originating from its unusual band structure and selection rules for the optical transitions near the Dirac point. Here we study the linear and nonlinear optical response of  graphene in strong magnetic and optical fields using quantum-mechanical density-matrix formalism. We calculate the power of coherent terahertz radiation generated as a result of four-wave mixing in graphene. We  show that even one monolayer of graphene gives rise to appreciable nonlinear frequency conversion efficiency and Raman gain for modest intensities of incident infrared radiation.

\end{abstract}

\maketitle

\section{Introduction and Background}

Graphene has unique electronic and optical properties stemming from linear, massless dispersion of electrons near the Dirac point and the chiral character of electron states\cite{castroneto,nair}. Magnetooptical properties of graphene and thin graphite layers are particularly interesting, showing multiple absorption peaks and unique selection rules for transitions between Landau levels \cite{sadowski,abergel2007,prl,kono2012}. Recent progress in growing high-quality epitaxial graphene and graphite with high room-temperature mobility and strong magnetooptical response attracted a lot of interest and showed the promise of new applications in the infrared optics and photonics \cite{orlita2008,orlita2009,crassee}.  The time is ripe to explore the nonlinear optical properties of a magnetized graphene and their applications. We have recently shown that graphene placed in a magnetic field possesses perhaps the highest infrared  optical nonlinearity among known materials \cite{prl}. Here we present detailed derivation of the linear and nonlinear response of a magnetized graphene based on a rigorous density-matrix formalism. We apply this approach to calculate the terahertz radiation power generated by third-order nonlinear optical processes: the four-wave mixing and stimulated Raman scattering. We argue that an extremely strong nonlinearity of graphene in combination with its unique selection rules makes graphene a promising material for the new generation of compact optoelectronic devices. 

.

\subsection{Band structure}

Graphene monolayer is a one-atom-thick monolayer of carbon atoms arranged in a hexagonal lattice, which we will treat as a perfect two-dimensional crystal structure in the ($x,y$)-plane. 
The electronic band structure of graphene has been extensively studied, starting from Wallace in 1947 \cite{Wallace} who used the tight-binding model. Here we briefly summarize the results  relevant for our subsequent derivation of the optical response. The carbon atom in graphene has four valence electrons, three of which form tight bonds with the three neighbor atoms, therefore one atom only has one conduction electron in $2 p_z$ state. Considering only the interaction with the atom's three nearest neighbors, the resulting Hamiltonian in  k-representation is purely off-diagonal: 
$$
\hat{H}_K=\left(\begin{array}{cc}
0 & -\gamma_0\e^{-ik_xa}\left[1+2\e^{+i\frac{3k_xa}{2}}\cos\sqrt{\frac{3}{2}}k_ya\right] \\
-\gamma_0\e^{+ik_xa}\left[1+2\e^{-i\frac{3k_xa}{2}}\cos\sqrt{\frac{3}{2}}k_ya\right] & 0
\end{array}\right)
$$
Here $\gamma_0 \sim 2.8$ eV and $a = 1.42$ $\AA$  are the nearest-neighbor hopping energy and C-C spacing. Then the energy dispersion relation is
\begin{equation}
E_k=\pm \gamma_0 \sqrt{ 1+4\cos\frac{\sqrt{3}k_xa}{2} \cos\frac{k_ya}{2}+4\cos^2\frac{k_ya}{2}}
\end{equation}

The electron and hole bands, denoted by $\pm$ are fully symmetric about the Dirac points at the six corners of the first Brillouin zone,  $$\frac{\sqrt{3}k_xa}{2}=(2n+1)\pi, \cos\frac{k_ya}{2}=0.5,$$ where $E_k=0$. Only two of these six Dirac points are inequivalent, referred to as K and K'. The  Dirac points are located on the Fermi level if graphene is undoped and unbiased.
In the vicinity of Dirac points, for example near $k=K$, it is convenient to define $\vec{q} = \vec{k}-\vec{K}$. The effective Hamiltonian then becomes
\begin{equation}
\hat{H}_q=\hbar\upsilon_F\left(\begin{array}{cc}
0 & q_x+iq_y \\
q_x-iq_y & 0 \\
\end{array}\right)=\hbar\upsilon_F\hat{\vec{\sigma}}\cdot\vec{q},
\end{equation}
similarly to the one for an ultra-relativistic massless particle with spin 1/2 after replacing the velocity of light $c$ with the band parameter (Fermi velocity)  $\upsilon_F = 3 \gamma_0/2\hbar a \sim 10^6$ m/s. The pseudospin variable entering the problem is related to the presence of two sublattices A and B \cite{castroneto}. The corresponding eigenvalue is $$E(\vec{q})= \pm\hbar\upsilon_F\sqrt{q_x^2+q_y^2},$$ and the eigenfunctions are two-component spinors.


\subsection{Landau levels}
In an external magnetic field $B \hat{z}$ perpendicular to the plane of graphene, the continuous energy bands near the Dirac points split into discrete Landau levels. The  effective-mass Hamiltonian \cite{Ando:05,Ando:02,Ando:07} for a graphene monolayer can be written as a 4x4 matrix to combine the contribution of K and K' points:  
\begin{equation}
\hat{H}_0=\upsilon_F\left(\begin{array}{cccc}
0 & \hat{\pi}_x-i\hat{\pi}_y & 0 & 0 \\
\hat{\pi}_x+i\hat{\pi}_y & 0 & 0 & 0 \\
0 & 0 & 0 & \hat{\pi}_x+i\hat{\pi}_y \\
0 & 0 & \hat{\pi}_x-i\hat{\pi}_y & 0
\end{array}\right)
\end{equation}
where $\hat{\vec{\pi}}=\hat{\vec{p}}+e\vec{A}/c$, $\hat{\vec{p}}$ the electron momentum operator, and $\vec{A}$ is the vector potential, which is equal to $(0,Bx)$ for a constant magnetic field.
 In this Hamiltonian the coupling between the K and K' point is neglected, so we can write down the solutions to the Schr\"{o}dinger equation $ \hat{H} \Psi =\varepsilon \Psi $
separately  for each point. For example, near the K point the Hamiltonian is  $\hat{H}_K=\upsilon_F \hat{\vec{\sigma}} \cdot \hat{\vec{\pi}}$ and the eigenfunction is specified by two quantum numbers $n$ and $k_y$, where $n=0,\pm1,\pm2,\cdots$, and $k_y$ is the electron wave vector along $y$ direction:
\begin{equation}
\Psi_{n,k_y}(r)=\frac{C_n}{\sqrt{L}}\exp(-ik_y y)\left(\begin{array}{c}
{\rm sgn}(n)i^{|n|-1}\phi_{|n|-1}\\
i^{|n|}\phi_{|n|}
\end{array}\right)
\end{equation}
with $$ C_n=\left\{\begin{array}{cc}
1 &(n=0) \\
\frac{1}{\sqrt{2}} &(n\ne0)
\end{array}\right.
$$
and
$$
\phi_{|n|}= \displaystyle \frac{H_{|n|}\left( \displaystyle (x-l_c^2k_y)/l_c\right)}{\sqrt{2^{|n|}|n|!\sqrt{\pi}l_c}}\exp{\left[-\frac{1}{2} \left(\frac{x-l_c^2k_y}{l_c}\right)^2\right]} ,
$$
where $l_c=\sqrt{c\hbar/eB}$ is magnetic length, $H_n(x)$ the Hermite polynomial.
The eigen energy is $$ \varepsilon_n={\rm sgn}(n)\hbar\omega_c \sqrt{|n|},\; \omega_c =\sqrt{2}\upsilon_F/l_c.$$
Positive or negative value of $n$ corresponds to electrons or holes. Compared with Landau levels for a conventional 2D electron/hole system with a parabolic dispersion, $E_n = (n+1/2)\hbar e B / m^*,$ Landau levels in graphene are unequally spaced: $\propto \sqrt{B}$. As shown in Fig.~1, the magnetic field "condenses" the original states in the Dirac cone into discrete energies, and each Landau level contains the same aerial density of states $N_{\Phi}=1/2\pi {l_c}^2$, not including spin and valley degeneracy factors.

\section{Optical transitions between the Landau levels}

\subsection{Selection rules}

Transitions between adjacent Landau levels in graphene fall into the mid-infrared to terahertz (THz) range for a magnetic field in the range 0.01-10 Tesla: $\hbar \omega_c \simeq 36 \sqrt{B({\rm Tesla})}$ meV.  Consider an incident classical optical field $\vec{E}=E(\omega)\exp{(-i\omega t)}\hat{e}$ polarized in the x-y plane along vector $\hat{e}$. Let us define the left-hand circular polarization vector as $\hat{e}_{LHS}=[\hat{x}-i\hat{y}]/\sqrt{2}$ and the right-hand circular polarization vector $\hat{e}_{RHS}=[\hat{x}+i\hat{y}]/\sqrt{2}$.  To include interaction with the optical field, we add its vector potential , $\vec{A}_{opt}=ic \vec{E}/\omega$, to the vector potential of the magnetic field in the generalized momentum operator $\hat{\vec{\pi}}$ in the Hamiltonian. This results in adding the interaction Hamiltonian $\hat{H}_{int}$ to $\hat{H}_0$, where
\begin{equation} \label{int}
\hat{H}_{int} = \upsilon_F \hat{\vec{\sigma}}\cdot \frac{e}{c}\vec{A}_{opt}
\end{equation}

Unlike the interaction Hamiltonian $H_{int}$ for an electron with a parabolic dispersion, there are no higher order terms such as $\pi^2$ near the Dirac point in graphene, so that even for a relatively strong optical field the interaction Hamiltonian is still linear with respect to $\vec{A}_{opt}$. Furthermore, $H_{int}$ does not contain the momentum operator; it is simply determined by the Pauli matrix vector $\hat{\vec{\sigma}}$. The matrix element of the optical transition between Landau levels is given by 
$$
\left< i|H_{int}|j\right> = \frac{i\upsilon_F}{\omega}\left< i |\sigma_x \hat{x} + \sigma_y \hat{y} | j\right> \cdot \vec{E} ,
$$
where $\left< i | \sigma_x \hat{x} + \sigma_y \hat{y} | j \right> $ is 
$$
\sqrt{2}C_i C_j(-i)^{|n_i|+|n_j|-1}\left({\rm sgn}(n_i) \left< \phi_{|n_i|-1}|\phi_{|n_j|}\right> \cdot \hat{e}_{LHS} + {\rm sgn}(n_j)\left<\phi_{|n_i|}|\phi_{|n_j|-1}\right> \cdot \hat{e}_{RHS} \right). 
$$

Since $\phi_n$ are orthogonal, the above expression is nonzero only when $|n_i|-1=|n_j|$ or $|n_i|=|n_j|-1$.
As a result, the selection rule for the allowed transitions turns out to be 
\begin{equation}\Delta |n| = \pm 1,
\end{equation} 
where n is the energy quantum number. Denoting $n_f$ and $n_i$ as the quantum numbers of the final and initial state, we can also conclude that $\hat{e}_{RHS}$ photons are absorbed when $|n_f| = |n_i| - 1$ while an absorption of a $\hat{e}_{LHS}$ photon leads to the transition $|n_f| = |n_i| + 1$. Comparing with a typical selection rule for inter-Landau level transitions in a traditional 2D system, $\Delta n = \pm 1$, the transitions with $\Delta n$  greater than 1 are allowed in graphene, for example, from $n_i=-1$ to $n_f=2$, which leads to an efficient resonant nonlinear mixing. Mid/far-infrared optical absorption between  Landau levels in monolayer and multilayer graphene has been extensively studied theoretically and in experiments; see e.g. \cite{sadowski,abergel2007,prl,kono2012,orlita2008,orlita2009,crassee}.

\subsection{The dipole moment matrix of graphene}

To calculate the 2D optical polarization as an average dipole moment per unit area of the graphene sheet,
\begin{equation}
 \vec{P}(\vec{r},t)=N \langle \vec{\mu} \rangle = N {\rm tr} (\hat{\rho} \cdot \hat{\vec{\mu}})\, ,
\end{equation}
where $N$ is the surface density of electrons and $\hat{\rho}$ is their density matrix, we need to know the dipole moment matrix $\hat{\mu}$ associated with inter-Landau level transitions. To calculate it, we first evaluate the commutator 
$$
[ \hat{\vec{r}} ,  \hat{H} ] = [ \hat{\vec{r}} , \upsilon_F \hat{\vec{\sigma}} \cdot \hat{\vec{p}}] + [ \hat{\vec{r}} , \upsilon_F \hat{\sigma} \cdot \frac{e}{c} \vec{A}] .
$$
The second term on the right-hand side is zero because $\vec{A}$ is a function of $\vec{r}$. So the commutator of $\hat{\vec{r}}$ and the Hamiltonian is
$$
[ \hat{\vec{r}} , \hat{H} ] = \upsilon_F \hat{\vec{\sigma}} \cdot [ \hat{\vec{r}} , \hat{\vec{p}} ] = i \hbar \upsilon_F \hat{\vec{\sigma}} .
$$
Since we use hats for both the operators and the unit vectors, we will stop putting hats over vector-valued operators unless it may create confusion. Choosing the eigen states of $\hat{H}$ as the basis, we obtain 
$$
\langle m | [ \vec{r} , \hat{H} ] | n \rangle = \langle m | \vec{r} \hat{H} | n \rangle - \langle m | \hat{H} \vec{r} | n \rangle = (\varepsilon_n - \varepsilon_m) \langle m | \vec{r} | n \rangle ,
$$
where $\varepsilon_n$ and $\varepsilon_m$ are the eigen energies of states $| n \rangle$ and $| m \rangle$. So the dipole matrix element of a closed system is defined as
\begin{equation}
\vec{\mu}_{mn} = e \cdot \langle m | \vec{r} | n \rangle = \frac{i\hbar e}{\varepsilon_n - \varepsilon_m} \langle m | \upsilon_F \hat{\vec{\sigma}} | n \rangle .
\end{equation}

Similarly to the  matrix elements of the interaction Hamiltonian $\langle m | \hat{H}_{int} | n \rangle$, the dipole matrix elements are determined by elements of the Pauli matrix  $\langle m | \hat{\sigma} | n \rangle$. In particular, $\vec{\mu}_{mn}$ is nonzero when $|m| = |n| \pm 1$. Using the wavefunction in Eq.~(4), the analytic expression for the dipole moment element can be derived:
\begin{eqnarray}
\vec{\mu}_{mn} & = & \frac{i\hbar e \upsilon_F}{\varepsilon_n - \varepsilon_m} C_m^* C_n \left( {\rm sgn}(m)(-i)^{|m|-1}\phi_{|m|-1}^* ,\, (-i)^{|m|}\phi_{|m|}^* \right)\nonumber \\
  & &\cdot  \left( \, \begin{array}{cc}
 0 & \hat{x}+i\hat{y} \, \\
 \hat{x}-i\hat{y} & 0 \,
 \end{array} \,\right) \cdot
 \left(\,\begin{array}{c}
 {\rm sgn}(n)i^{|n|-1}\phi_{|n|-1} \\
 i^{|n|-1}\phi_{|n|}
 \end{array} \, \right) \nonumber \\
 & = & \frac{i\hbar e \upsilon_F C_m C_n (-1)^{|m|-1}i^{|m|+|n|-1}}{\varepsilon_n - \varepsilon_m} \left( {\rm sgn(m)}\delta_{|m|-1,|n|}(\hat{x}-i\hat{y})\right.\nonumber \\
 &&\left.- {\rm sgn(n)}\delta_{|m|,|n|-1}(\hat{x}+i\hat{y})\right) \label{mu}
\end{eqnarray}

As an example, in a 4-level system that will be considered below (energy quantum numbers n= -1, 0, 1, 2), the allowed transitions are between $n=\pm 1$ and $n = 0$, and between $n=\pm 1$ and $n=2$. The eigen functions of these four energy levels are
\begin{eqnarray} \displaystyle 
| 1 \rangle & = & \frac{1}{\sqrt{2L}}\exp(-ik_{y1} y)\left(\begin{array}{c}
- \phi_{0}\\
i\phi_{1}
\end{array}\right) \nonumber \\
| 2 \rangle & = & \frac{1}{\sqrt{L}}\exp(-ik_{y2} y)\left(\begin{array}{c}
0\\
\phi_{0}
\end{array}\right) \nonumber \\
| 3 \rangle & = & \frac{1}{\sqrt{2L}}\exp(-ik_{y3} y)\left(\begin{array}{c}
 \phi_{0}\\
i\phi_{1}
\end{array}\right) \nonumber \\
| 4 \rangle & = & \frac{1}{\sqrt{L}}\exp(-ik_{y4} y)\left(\begin{array}{c}
i \phi_{1}\\
- \phi_{2}
\end{array}\right)
\end{eqnarray}
Combining with Eq.~(\ref{mu}), the dipole moment matrix of the 4-level system is
\begin{equation}
\vec{\mu}= \displaystyle  \frac{e\upsilon_F}{\sqrt{2} \omega_c}
\left(
\begin{array}{cccc}
 0&-i\hat{x}-\hat{y}&0&\frac{i\hat{x}-\hat{y}}{2+\sqrt{2}}  \\
 i\hat{x}-\hat{y}&0&i\hat{x}-\hat{y}&0 \\
 0&-i\hat{x}-\hat{y}&0&\frac{i\hat{x}-\hat{y}}{2-\sqrt{2}}  \\
 \frac{-i\hat{x}-\hat{y}}{2+\sqrt{2}}&0&\frac{-i\hat{x}-\hat{y}}{2-\sqrt{2}}&0 
\end{array}\right)
\end{equation}

To summarize, the dipole moment of the transition between the Landau levels in graphene has a magnitude of the order of 
$$|\mu_{mn}| \sim \displaystyle \frac{e \hbar \upsilon_F}{\varepsilon_n - \varepsilon_m} \propto 1/\sqrt{B}.$$ 
This is a very large value for the transitions near the Dirac point where $\varepsilon_n - \varepsilon_m \sim \hbar \omega_c$: $\upsilon_F/\omega_c \sim 18$ nm at $B = 1$ T. Note that the dipole moment grows rapidly, $\sim \lambda$, with increasing transition wavelength. This is a faster growth than in atomic systems ($\sim \sqrt{\lambda}$)  or conventional semiconductors. Therefore, one expects a very strong nonlinear optical response in the mid-infrared and THz region.


\section{Linear optical response of graphene in a magnetic field}

\subsection{Linear susceptibility of graphene}

The optical polarization of the graphene sheet,
$ \vec{P}(\vec{r},t) = N {\rm tr} (\hat{\rho} \cdot \hat{\vec{\mu}})$, 
can be presented as a usual expansion in terms of electric susceptibilities if the density matrix is solved as a series in powers of the incident fields. 
For a weak monochromatic field it is enough to keep the first, linear in $E$, term in the expansion of the density matrix to find the linear susceptibility $\chi^{(1)}$. 

The equation of motion for the density matrix elements is given by 
\begin{equation} \label{rho}
 \dot{\rho}_{nm}=-\frac{i}{\hbar}(\varepsilon_n - \varepsilon_m)\rho_{nm}-\frac{i}{\hbar}[ \hat{H}_{int}(t), \hat{\rho} ]_{nm}-\gamma_{nm}(\rho_{nm}-\rho_{nm}^{(eq)}).
\end{equation}
Here we approximated incoherent scattering with phenomenological decay rates $\gamma_{nm}$ describing the relaxation of the matrix elements to their equilibrium values $\rho_{nm}^{(eq)}$; $\rho_{nm}^{(eq)} = 0$ for $n \neq m$. 

Formal expansion of the density matrix in powers of the interaction Hamiltonian leads to the following differential equations:
$$
\dot{\rho}_{nm}^{(0)} = -i \omega_{nm} \rho_{nm}^{(0)} - \gamma_{nm} (\rho_{nm}^{(0)} - \rho_{nm}^{(eq)}) \,;
$$
$$
\dot{\rho}_{nm}^{(1)} = -(i \omega_{nm} + \gamma_{nm}) \rho_{nm}^{(1)} - \frac{i}{\hbar} [\hat{H}_{int}, \hat{\rho}^{(0)}]_{nm} \,;
$$
$$
\dot{\rho}_{nm}^{(2)} = -(i \omega_{nm} + \gamma_{nm}) \rho_{nm}^{(2)} - \frac{i}{\hbar} [\hat{H}_{int}, \hat{\rho}^{(1)}]_{nm} \,;
$$
$$
\dots \dots \dots
$$
where $\omega_{nm} = (\varepsilon_n - \varepsilon_m)/\hbar$. 
Choosing $\rho_{nm}^{(0)} = \rho_{nm}^{(eq)}$, we can calculate higher order terms step by step. The iteration formula is given by
\begin{equation} 
\rho_{nm}^{(N)} = \int^{t} \frac{-i}{\hbar} [\hat{H}_{int}(t'), \hat{\rho}^{(N-1)}]_{nm} \exp{[(i \omega_{nm} + \gamma_{nm})\cdot (t'-t)]}dt' . \label{iter}
\end{equation}

The interaction Hamiltonian was derived above. It can be rewritten as
\begin{eqnarray}
\hat{H}_{int}(t)&=&\upsilon_F \vec{\sigma} \cdot \frac{e}{c} \vec{A}_{opt} \nonumber \\
&=&-\frac{ie\upsilon_F}{\omega}\vec{\sigma} \cdot \vec{E}(\omega)\exp{(-i\omega t)} \nonumber \\
& \equiv & -\tilde{\vec{\mu}}\cdot \vec{E}(\omega)\exp{(-i\omega t)} \label{hint} 
\end{eqnarray}
Here we have defined
$$\tilde{\vec{\mu}} \equiv \frac{ie\upsilon_F}{\omega}\vec{\sigma};\; \langle m|\tilde{\vec{\mu}}|n \rangle\equiv \frac{ie\upsilon_F}{\omega}\langle m|\vec{\sigma}|n \rangle, $$
which coincides with the dipole moment if the incident optical field is exactly on resonance with a given transition, that is $\tilde{\vec{\mu}}_{mn}=\vec{\mu}_{mn} $ when $\varepsilon_n - \varepsilon_m = \hbar \omega$.

The first-order (linear) part of the density matrix can then be calculated from the iteration formula Eq.~(\ref{iter}) and Eq.~(\ref{hint}):
\begin{equation}
\rho_{nm}^{(1)} = \int_0^{t} \frac{-i}{\hbar} [\hat{H}_{int}(t'), \hat{\rho}^{(eq)}]_{nm} \exp{[(i \omega_{nm} + \gamma_{nm})\cdot (t'-t)]}dt' ,
\end{equation}
where
\begin{eqnarray}
[\hat{H}_{int}(t'), \hat{\rho}^{(eq)}]_{nm} & = & \sum_{\upsilon}\left( \tilde{\vec{\mu}}_{n\upsilon}\rho_{\upsilon m}^{(eq)}-\rho_{n\upsilon}^{(eq)}\tilde{\vec{\mu}}_{\upsilon m}\right) \cdot \tilde{E}(t') \nonumber \\
& = & \left(\rho_{mm}^{(eq)}-\rho_{nn}^{(eq)}\right)\tilde{\vec{\mu}}_{nm}\cdot \tilde{E}(t') .
\end{eqnarray}
This yields the 2D first-order polarization  in the form 
\begin{eqnarray}
\tilde{P}^{(1)}(\omega) &=& N {\rm tr} \left(\hat{\rho}^{(1)}\hat{\mu}\right) \nonumber \\
&=& N \sum_{nm} \frac{\rho_{mm}^{(eq)}-\rho_{nn}^{(eq)}}{\hbar}\cdot\frac{\left(\tilde{\vec{\mu}}_{nm}\cdot\hat{e}\right)\vec{\mu}_{mn}}{(\omega_{nm}-\omega)-i\gamma_{nm}}
E(\omega)\exp{(-i\omega t)} .
\end{eqnarray}
Here $N$ is the 2D (sheet) electron density of graphene, which is $n_s n_{\upsilon} N_{\Phi} = 2/(\pi l_c^2)$ where $n_s=2$ and $n_{\upsilon}=2$ are spin and valley degeneracy. For a left-hand polarized optical field, the circular polarization vector $\hat{e}$ is $\hat{e}_{LHS}=[\hat{x}-i\hat{y}]/\sqrt{2}$, and the term $\left(\tilde{\vec{\mu}}_{nm}\cdot\hat{e}\right)$ in the above expression is nonzero only when $|n|$=$|m|-1$. On the other hand, for a right-hand polarized optical field the term is nonzero only when $|m|$=$|n|-1$. This of course corresponds to the polarization selection rules that were already derived above; see also \cite{abergel2007,prl,kono2012}. Taking them into account, the expressions for the 2D linear optical susceptibility for the left/right-hand in-plane polarized optical field are:
\begin{eqnarray}
\chi^{(1)}(\omega, \hat{e}_{LHS})&=& \sum_{|n|=|m|-1}\frac{-4C^2_m C^2_n e^2\upsilon_F^2}{\pi l_c^2 \hbar \omega \omega_{nm}}\cdot \frac{\rho_{mm}^{(eq)}-\rho_{nn}^{(eq)}}{\omega_{nm}-\omega-i\gamma_{nm}} \nonumber \\
\chi^{(1)}(\omega, \hat{e}_{RHS})&=& \sum_{|m|=|n|-1}\frac{-4C^2_m C^2_n e^2\upsilon_F^2}{\pi l_c^2 \hbar \omega \omega_{nm}}\cdot \frac{\rho_{mm}^{(eq)}-\rho_{nn}^{(eq)}}{\omega_{nm}-\omega-i\gamma_{nm}}
\label{chi1} 
\end{eqnarray}

\subsection{Absorption coefficient}

The high-frequency absorbance in monolayer graphene at zero magnetic field, $\alpha =\pi e^2 / \hbar c$, is a constant. In a high magnetic field, the absorption coefficient shows a series of peaks due to inter-Landau-level transitions. From the standard expression for a weak absorption, 
\begin{equation}
\alpha \simeq  \frac{4\pi\omega}{c}{\rm Im}[\chi^{(1)}(\omega)],
\end{equation}
and combining with Eq.~(\ref{chi1}),  we can calculate the absorption coefficient of monolayer graphene for the left/right-hand in-plane polarized optical field:

\begin{eqnarray}
\alpha(\omega, \hat{e}_{LHS})&=& \sum_{|n|=|m|-1}\frac{-16C^2_m C^2_n e^2\upsilon_F^2 \gamma_{nm}}{ l_c^2 \hbar c\omega_{nm}}\cdot \frac{\rho_{mm}^{(eq)}-\rho_{nn}^{(eq)}}{(\omega_{nm}-\omega)^2+\gamma^2_{nm}} \nonumber \\
\alpha(\omega, \hat{e}_{RHS})&=& \sum_{|m|=|n|-1}\frac{-16C^2_m C^2_n e^2\upsilon_F^2 \gamma_{nm}}{ l_c^2 \hbar c \omega_{nm}}\cdot \frac{\rho_{mm}^{(eq)}-\rho_{nn}^{(eq)}}{(\omega_{nm}-\omega)^2+\gamma^2_{nm}} \label{abs} 
\end{eqnarray}
The transition linewidth is of the order of 10 meV; for example, it was measured to be $\sim 30$ meV in the magnetic field of 3 T \cite{Kim}. The corresponding relaxation rate $\gamma$ is then on the scale of a few $\sim 10^{13}$ s$^{-1}$. However, we should keep in mind that this number depends on the sample quality and the substrate used in the experiment.

The above result agrees with the absorption coefficient calculated in \cite{abergel2007} using the Keldysh's Green function approach. If we assume the relaxation rates between different levels to be the same, that is $\gamma_{nm}=\gamma$,  and follow their notation for Landau levels as $(\alpha, n)$, where $n\geq 0$ and $\alpha = \pm 1$ denote whether the corresponding state is in conduction (+) or valence (-) band, Eq.~(\ref{abs}) can be rewritten in the form identical to the one in \cite{abergel2007}:
\begin{eqnarray}
\alpha(\omega, \hat{e}_{LHS})&=& \sum_{n\geq0, \alpha \alpha'}\frac{\frac{8C^2_n\upsilon^2_F e^2 \gamma}{l^2_c \hbar c(\alpha\sqrt{n+1}\omega_c-\alpha'\sqrt{n}\omega_c)}(\rho_{n,\alpha'}-\rho_{n+1,\alpha})}{(\alpha\sqrt{n+1}\omega_c-\alpha'\sqrt{n}\omega_c-\omega)^2+\gamma^2}, \nonumber \\
\alpha(\omega, \hat{e}_{RHS})&=& \sum_{n\geq1, \alpha \alpha'}\frac{\frac{8C^2_{n-1}\upsilon^2_F e^2 \gamma}{l^2_c \hbar c(\alpha\sqrt{n-1}\omega_c-\alpha'\sqrt{n}\omega_c)}(\rho_{n,\alpha'}-\rho_{n-1,\alpha})}{(\alpha\sqrt{n-1}\omega_c-\alpha'\sqrt{n}\omega_c-\omega)^2+\gamma^2}.
\end{eqnarray}

\section{Nonlinear optical response}

Strong optical nonlinearity of graphene, like most of its unique electrical and optical properties, stems from the peculiar energy dispersion of carriers near the Dirac points, $E \propto \pm |\vec{p}|$. As a result, the electron velocity $\propto \partial E/\partial \vec{p}$  induced by an incident  electromagnetic wave is a strongly nonlinear function of induced electron momentum. Nonlinear electromagnetic response of classical charges with such an energy dispersion has been studied theoretically in \cite{mikhailov2008}. Recently, the four-wave mixing in graphene without a magnetic field has been observed at near-infrared wavelengths
\cite{hendry2010}. Effective bulk third-order susceptibility was estimated to have a very large value, $\chi^{(3)} \sim 10^{-7}$ esu, which is more than an order of magnitude larger than in gold films. 

Nonlinear cyclotron resonance in graphene was considered theoretically  in \cite{mikhailov2009}, again in the classical limit, which can be applied only to electrons in a low magnetic field that occupy highly excited Landau levels $n \gg 1$, when energy and momentum quantization are neglected. In a recent work \cite{prl} we presented a quantum mechanical density-matrix description of the nonlinear optical response of graphene, which is valid for quantizing magnetic fields and strong optical fields, including the effect of saturation of inter-Landau level transitions. Due to unique optical selection rules for "massless" electrons near the Dirac point, one can implement a nonlinear interaction in which all optical fields are resonant to allowed optical transitions. The resulting magnitude of $\chi^{(3)}$ turns out to be extremely large, of the order of $0.1$ esu at mid/far-infrared wavelengths in the field of several Tesla. A similar strategy of a completely resonant nonlinear wave mixing has been implemented in asymmetric coupled quantum well systems, where one can increase the dipole moment of an intersubband transition involving a large change in the energy quantum number $n$ by an appropriate band structure design \cite{gurnick,sirtori,gmachl,malis,mosely2004,belyanin2005,faist}. However, the resulting third-order nonlinearity was still several orders of magnitude lower than in graphene for the same spectral range. 

\subsection{Four-wave mixing}

Efficient nonlinear optical coupling becomes possible in graphene due to strong non-equidistance of the Landau levels, large magnitude of the dipole matrix elements, and unusual selection rules $\Delta |n| = \pm 1$ which enable transitions with change in $n$ greater than 1. In this section we study a specific example of the nonlinear optical interaction, namely the four-wave mixing. Consider a strong bichromatic field $\vec{E} = \vec{E}_1\exp(-i\omega_1 t) + \vec{E}_2 \exp(-i\omega_2 t) + {\rm c.c.}$ normally incident on the graphene layer. Here $\omega_1$ is nearly resonant with the transition from $n = -1$ to $n = 2$ and $\vec{E}_1$ has left circular polarization. The frequency $\omega_2$ is nearly resonant with the transition from $n = 0$ to $n = \pm 1$ and $\vec{E}_2$ has linear polarization, so that it couples both to transition $-1 \rightarrow 0$ and $0 \rightarrow 1$, as shown in Fig.~1. As a result of the four-wave mixing interaction, the right-circularly polarized field $\vec{E}_3$ at frequency $\omega_3 = \omega_1 - 2 \omega_2$ nearly resonant with the transition from $n = 2$ to $n = 1$ is generated.

\begin{figure}[htb]
\centerline{
\includegraphics[width=10cm]{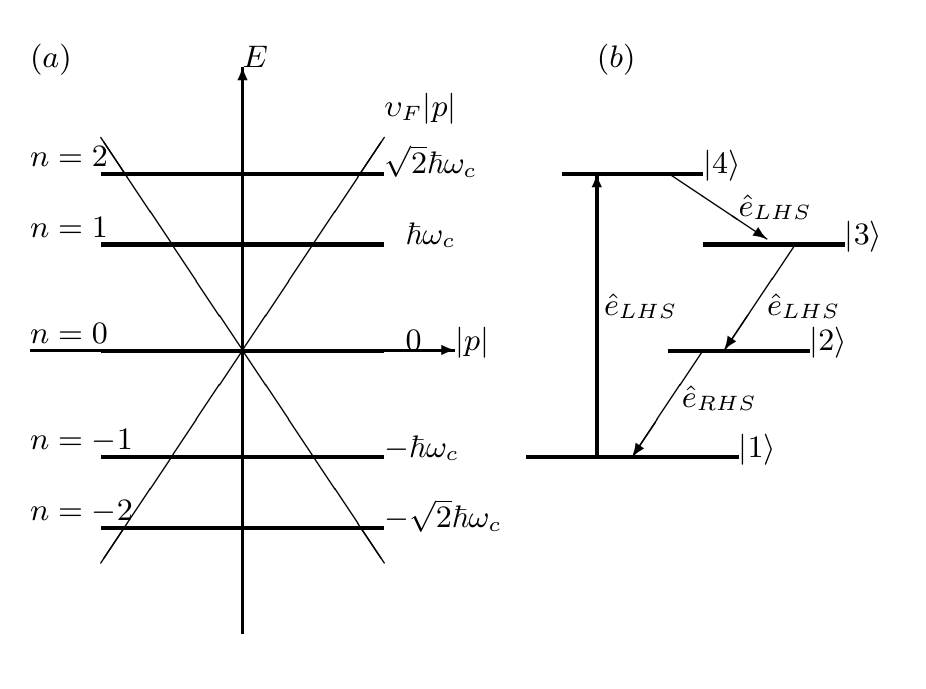}}
 \caption{Landau levels near the Dirac point superimposed on the electron dispersion without the magnetic field $E=\pm\upsilon_F |p|$. (b): A scheme of the four-wave mixing process in the four-level system of Landau levels with energy quantum numbers $n = -1, 0, +1, +2$ that are renamed to states 1 through 4 for convenience. }
\end{figure}

The frequencies involved in the four-wave mixing fall into the mid-infrared and THz region in the magnetic field of a few Tesla, as shown in Fig.~2. For example, at $B = 3$T, the nonlinear signal is generated at a wavelength of 48 $\mu$m in the presence of pump fields at wavelengths 8 and 20 $\mu$m.
\begin{figure}[htb]
\centerline{
\includegraphics[width=8cm]{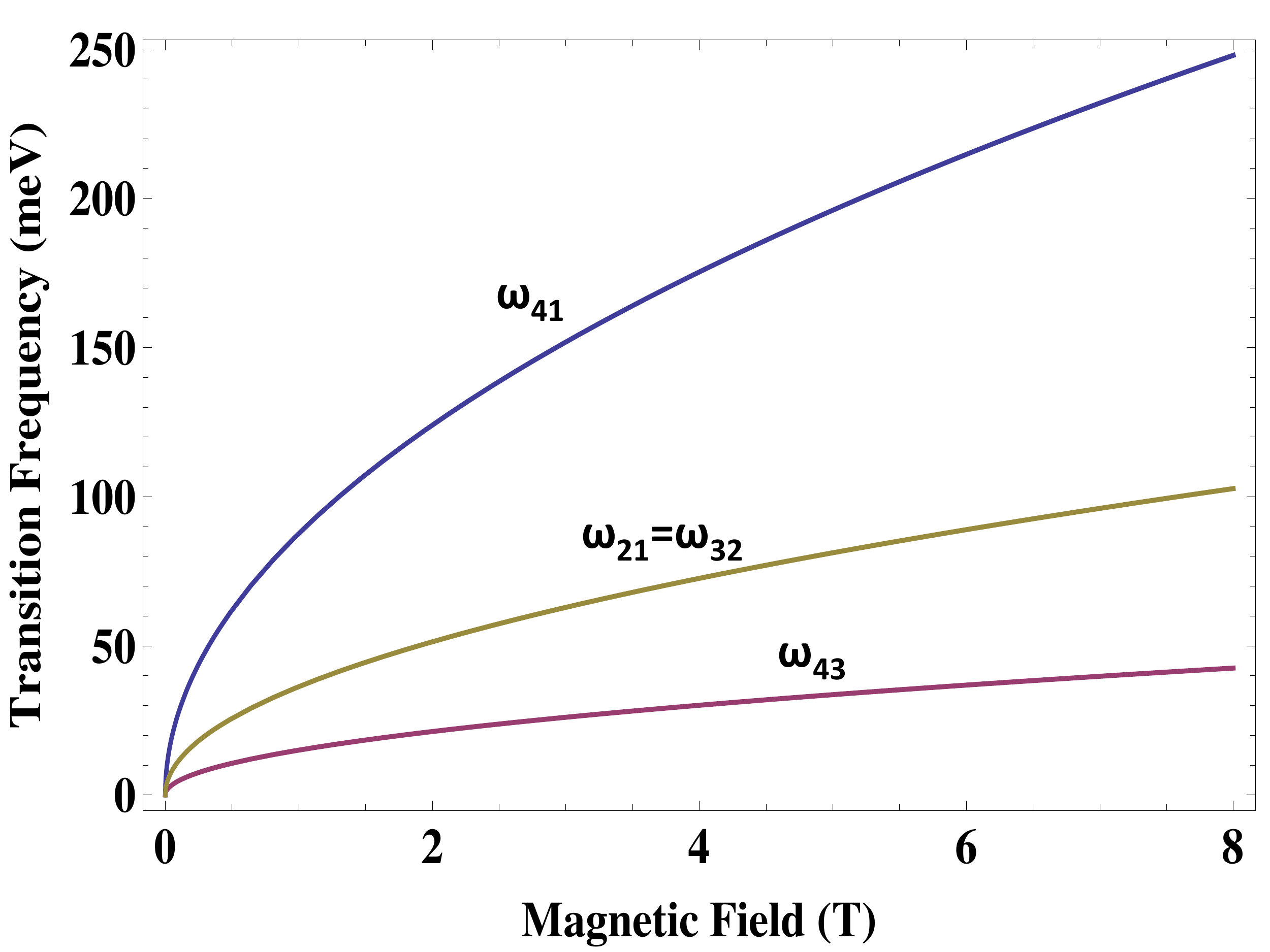}}
 \caption{Transition frequencies in the 4-level graphene system shown in Fig.~ 1(b). $\omega_{mn}$ indicates the transition frequency between levels $m$ and $n$. }
\end{figure}

Truncating the master equation (\ref{rho}) to the 4-level system shown in Fig.~1(b) and introducing slowly varying off-diagonal elements of the density matrix as  $\rho_{41} = \sigma_{41} e^{-i\omega_{1}t}$, $\rho_{41} = \sigma_{43} e^{-i\omega_{3}t}$, and $\rho_{32,21} = \sigma_{32,21} e^{-i\omega_{2}t}$,one can obtain the following set of equations for the amplitudes $\sigma_{nm}$ in the steady state and in the rotating wave approximation:

\begin{eqnarray}
i \Gamma_{21} \sigma_{21} & = & \Omega_{21}  n_{21} - \Omega^*_{32}\sigma_{31} + \Omega_{41}\sigma_{24} \nonumber\\
i \Gamma_{32} \sigma_{32} & = & \Omega_{32}  n_{32} - \Omega^*_{43}\sigma_{42} + \Omega^*_{21}\sigma_{31} \nonumber\\
i \Gamma_{43} \sigma_{43} & = & \Omega_{43}  n_{43} - \Omega_{41}\sigma_{13} + \Omega^*_{32}\sigma_{42} \nonumber\\
i \Gamma_{31} \sigma_{31} & = & \Omega_{21}  \sigma_{32} + \Omega_{41}\sigma_{34} - \Omega_{32}\sigma_{21} - \Omega^*_{43}\sigma_{41} \nonumber\\
i \Gamma_{42} \sigma_{42} & = & \Omega^*_{21} \sigma_{41} + \Omega_{32}\sigma_{43} - \Omega_{43}\sigma_{32} - \Omega_{41}\sigma_{12} \nonumber\\
i \Gamma_{41} \sigma_{41} & = & \Omega_{41}  n_{41} + \Omega_{21}\sigma_{42} - \Omega_{43}\sigma_{31}. \label{4level} 
\end{eqnarray}
Here the Rabi frequencies are defined as $\Omega_{ij} = \vec{E}_{ij}\cdot\tilde{\vec{\mu}}_{ij}/\hbar $ and the population differences are $n_{ij} = \rho_{ii}-\rho_{jj}$. The notation for the field amplitudes is as follows: $\vec{E}_{41} = \vec{E}_1$, $\vec{E}_{21}$ is the right circularly polarized component of $\vec{E}_2$, and $\vec{E}_{32}$ is the left circularly polarized component of $\vec{E}_2$. The complex dephasing  $\Gamma_{41} = \gamma_{41} + i(\omega_{41} - \omega_1)$ and similarly for other transitions; all detunings from resonance are small.  

If the incident field is not strong enough to perturb the populations,  the population differences $n_{mn}$ are constant in Eqs.~(\ref{4level}). As a result, the off-diagonal density matrix elements such as $\sigma_{43}$ can be solved analytically and written as an expansion in powers of the pump fields:
\begin{eqnarray}
\sigma_{43}  &=&  \frac{\Omega_{43}}{i\Gamma_{43}}n_{43} - \frac{\Omega_{41}\Omega^*_{21}\Omega^*_{32}}{i^3\Gamma_{43}\Gamma^*_{31}\Gamma^*_{32}} n_{32} +  \frac{\Omega_{41}\Omega^*_{21}\Omega^*_{32}}{i^3\Gamma_{43}\Gamma^*_{31}\Gamma^*_{21}} n_{21}
 + \frac{\Omega_{41}\Omega^*_{21}\Omega^*_{32}}{i^3\Gamma_{43}\Gamma_{42}\Gamma_{41}} n_{41}\nonumber \\
   &+&  \frac{\Omega_{41}\Omega^*_{21}\Omega^*_{32}}{i^3\Gamma_{43}\Gamma_{42}\Gamma^*_{21}} n_{21}
 +  \frac{|\Omega_{41}|^2\Omega_{43}}
{i^3\Gamma_{43}\Gamma^*_{31}\Gamma^*_{41}} n_{41} + ... \label{s43}
\end{eqnarray}
The first term on the right-hand side describes the linear absorption and the next four terms describe the 3rd order nonlinear optical response; the higher-order terms are dropped. Note that the last term on the right-hand side corresponds to a stimulated Raman scattering of the pump field $E_{41}$ into the signal field $E_{43}$, which we consider in the next section. 

The optical polarization at the frequency $\omega_3$ of the nonlinear signal in the rotating wave approximation is given by 
$$
\vec{P}(\omega_3) = N \cdot \sigma_{43}\vec{\mu}_{43}e^{-i\omega_{3}t} + c.c.
$$
Below we investigate different nonlinear effects contained in Eq.~(\ref{s43}). Consider first the four-wave mixing interaction $\omega_1 - 2\omega_2 \Rightarrow \omega_3$, described by the second through fourth terms on the right hand side of Eq.~(\ref{s43}). 

Substituting the expression for $\sigma_{43}$ into $\vec{P}(\omega_{3})$, and keeping only these three terms will lead to the third-order nonlinear susceptibility corresponding to the four-wave mixing:
\begin{eqnarray}
\chi^{(3)}(\omega_3) &=& \frac{N\mu_{43}\tilde{\mu}_{41}\tilde{\mu}^*_{32}\tilde{\mu}^*_{21}}{(i\hbar)^3\Gamma_{43}}\left( \frac{\rho_{22}-\rho_{33}}{\Gamma^*_{31}\Gamma^*_{32}} \right. \nonumber \\ &+&  \left.  \frac{\rho_{22}-\rho_{11}}{\Gamma^*_{31}\Gamma^*_{21}}-
\frac{\rho_{11}-\rho_{44}}{\Gamma_{42}\Gamma_{41}}+\frac{\rho_{22}-\rho_{11}}{\Gamma_{42}\Gamma^*_{21}}\right)
\label{chi3}
\end{eqnarray}

To estimate the order of magnitude of $\chi^{(3)}$, we assume that all incident fields are in exact resonance so that the detuning factors $\Gamma_{ij} = \gamma_{ij} = \gamma$ are real numbers and all dephasing rates are the same. We also assume for definiteness that state 1 is fully occupied while states 2, 3 and 4 are empty, which means $\rho_{11} = 1, \rho_{22} = \rho_{33} = \rho_{44} = 0$. Then the expression for $\chi^{(3)}$ is further simplified into 
\begin{equation}
\chi^{(3)}(\omega_3) \sim \frac{3N \mu_{43} \tilde{\mu}_{41} \tilde{\mu}_{32} \tilde{\mu}_{21}}{ (\hbar \gamma)^3}.
\label{chi3est}
\end{equation}

This expression contains a 2D electron density $N$ and is a 2D (surface) susceptibility. To convert it into the bulk susceptibility for comparison with other materials, we can divide it by the thickness of one monolayer $\Delta z \sim 3$ $\AA$.  Taking a reasonable value for the dephasing rate, $\gamma = 3\times 10^{13}$ s$^{-1}$ \cite{Kim},  the bulk weak-field susceptibility $\chi_{3D}^{(3)} \sim 0.37\, (1/B(T))$ esu = $ 5\times 10^{-9}\, (1/B(T))$ m$^2$/V$^2$. Here the magnetic field is measured in Tesla. This is by far the strongest nonlinearity as compared to any material we know.  

When the incident fields increase in intensity, they start affecting populations on each level. In this case Eqs.~(\ref{4level}) have to be solved together with the equations for diagonal components of the density matrix. Introducing phenomenological transition times $T_{ij}$ between levels $i$ and $j$, we can write these equations as $$
\frac{d n_1}{d t}  =  \frac{n_2}{T_{21}} + \frac{n_3}{T_{31}} + \frac{n_4}{T_{41}} - i\left( \Omega_{21}\sigma_{12} - \Omega_{12}\sigma_{21} + \Omega_{41}\sigma_{14} - \Omega_{14}\sigma_{41} \right)
$$
$$
\frac{d n_2}{d t}  =  \frac{n_3}{T_{32}} + \frac{n_4}{T_{42}} - \frac{n_2}{T_{21}} - i\left( \Omega_{12}\sigma_{21} - \Omega_{21}\sigma_{12} + \Omega_{32}\sigma_{23} - \Omega_{23}\sigma_{32} \right)
$$
$$
\frac{d n_3}{d t}  =  \frac{n_4}{T_{43}} - \frac{n_3}{T_{32}} - \frac{n_3}{T_{31}} - i\left( \Omega_{23}\sigma_{32} - \Omega_{32}\sigma_{23} + \Omega_{43}\sigma_{34} - \Omega_{34}\sigma_{43} \right)
$$
\begin{equation}
n_1  +  n_2 + n_3 + n_4 = 1 .
\end{equation}
This is of course a very crude approximation of the actual relaxation dynamics of electrons, but it allows us to estimate how the effects of the optical pumping and saturation affect the nonlinear mixing efficiency and power. 

 It is convenient to normalize incident fields by their saturation values which determine the field strength at which the population at a given transition becomes significantly perturbed:  
 \begin{equation}
E^s_{21} = \frac{\hbar \sqrt{\gamma_{21}/T_{21}}}{\mu_{21}};\quad E^s_{32} = \frac{\hbar \sqrt{\gamma_{32}/T_{32}}} {\mu_{32}};\quad E^s_{41} = \frac{\hbar \sqrt{\gamma_{41}/T_{41}}}{\mu_{41}}.
\end{equation}
Then the corresponding saturation Rabi frequencies are given by $\Omega_{nm} = \sqrt{\gamma_{nm}/T_{nm}}$. 
We then introduce the dimensionless fields $x$, $x'$, and $y$ as 
\begin{equation}
x = \Omega_{21}/\Omega^s_{21} ;\quad  x' = \Omega_{32}/\Omega^s_{32} ;\quad  y = \Omega_{41} / \Omega^s_{41}.
\end{equation}

For estimation, we can simply assume that the relaxation rates $T_{nm}$ are the same, $T_{nm} \sim T$. Then the solution to the density matrix equation of motion depends on the fields through only two dimensionless factors $x$ and $y$. In particular, the scaling Eq.~(\ref{chi3est})  for $\chi^{(3)}(\omega_3)$ becomes
\begin{equation}
\chi^{(3)}(\omega_3) \sim \frac{N \mu_{43} \tilde{\mu}_{41} \tilde{\mu}_{32} \tilde{\mu}_{21}}{ (\hbar \gamma)^3} \times f(x, y),
\end{equation}
where $f(x, y)$  is a function of $x$ and $y$ shown in Fig.~3. It is equal to 3 when incident fields are weak, $x,y \ll 1$, and quickly decreases as $x$ and $y$ become greater than one. 

The electric field of the generated signal is determined by the nonlinear polarization $\vec{P}^{(3)}(\omega_3)$. From Maxwell's equations, neglecting the depletion of the pump fields, we can obtain
\begin{equation} \label{maxw} \frac{\partial \vec{E}}{\partial z} = i \cdot \frac{2 \pi \omega}{c} \cdot \vec{P}. 
\end{equation}
Note that here $\vec{P}$ is a 3D polarization (an average dipole moment per unit volume).
\begin{figure}[htb]
\centerline{
\includegraphics[width=15cm]{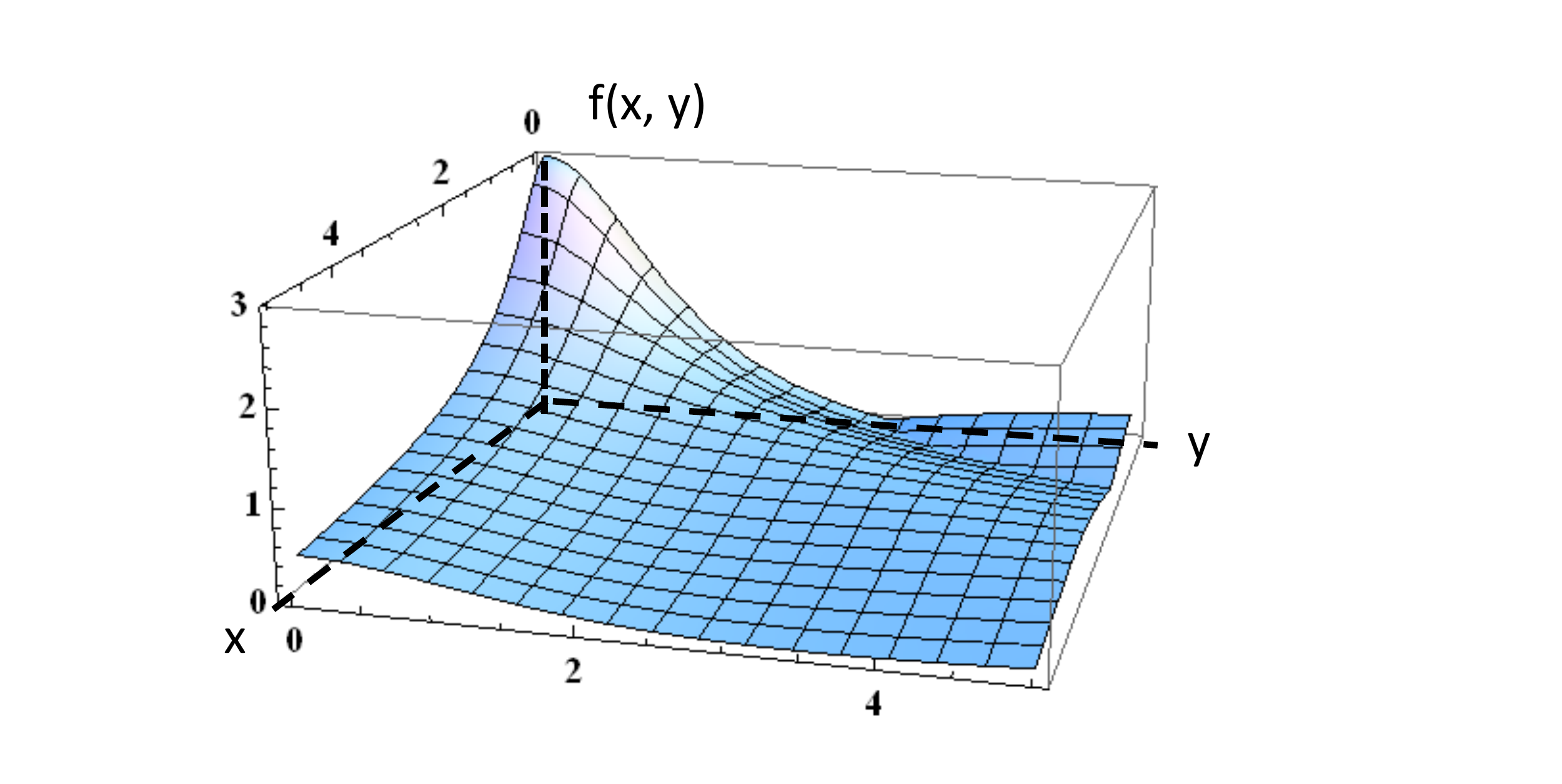}}
 \caption{Contour plot of $f(x,y)$ as a function of normalized pump fields $x$ and $y$. }
\end{figure}

For a thin layer of graphene one can integrate Eq.~(\ref{maxw}) over the thickness of the layer and obtain  
\begin{equation} \label{e3} 
E_3(\omega_3) = i \cdot \frac{2 \pi \omega_3}{c} \chi^{(3)}(\omega_3) E_1 (E^*_2)^2,
\end{equation}
where $\chi^{(3)}$ is a 2D susceptibility. The magnitude of $|\vec{E}_3|$ grows with the pump at small pump intensities and decays at high intensities because of the decrease in $\chi^{(3)}$. It reaches a maximum at $x =2.6$, $y=1.56$. Of course, these particular numbers depend on the relative values of the relaxation times between the Landau levels. However, the general conclusion that the maximum nonlinear signal is reached when the pump fields are of the order of the saturation values remains true. For fixed $x,y \sim 1$, $\chi^{(3)}$ scales with the magnetic field as $B^{-1}$, whereas $E \sim E_{sat} \sim \sqrt{B}$. As a result, from Eq.~(\ref{e3}), the maximum nonlinear signal scales with the magnetic field as 
\begin{equation} \label{scale}
|E^{max}_3| \sim   \omega_3 \chi^{(3)}(\omega_3) |E_{sat}|^3 \sim  \sqrt{B} \frac{1}{B} \cdot (\sqrt{B})^3 \sim B. 
\end{equation}
If we define   intensity as $I = c |E|^2 / 8\pi$, the intensity of the generated signal is related to the incident field intensities as
\begin{equation}
I_3(\omega_3)= \left( \frac{16 \pi^2 \omega_3}{c^2} \right)^2 |\chi^{(3)} |^2 I_1(\omega_1) (I_2(\omega_2))^2.
\end{equation}

Fig.~4(a) shows the plot of  $I_3$ as a function of the pump intensity $I_2$ when the second pump intensity $I_1$ is tied to $I_2$ by the optimal condition  $y = (1.56/2.6)x = 0.6x$.  The conversion efficiency in the magnetic field of 1-10 T is $I_3/I_2 \sim 10^{-5} - 10^{-6}$. 
\begin{figure}[htb]
\centerline{
\includegraphics[width=15cm]{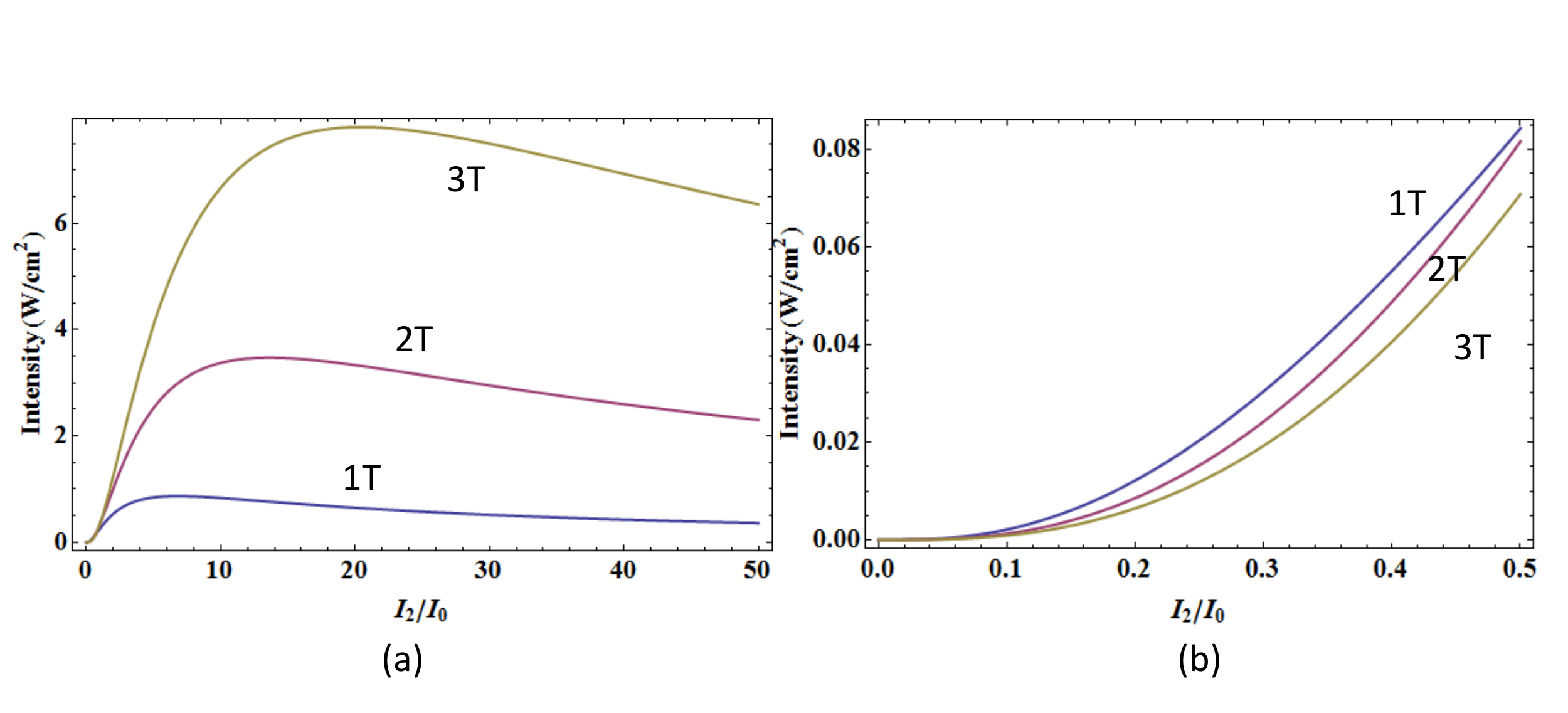}}
 \caption{(a) Intensity of the 4-wave mixing signal as a function of the intensity of the pump field $E_2$ normalized by $I_0 = c|E_{\rm sat}|^2/8\pi \simeq 2.2\times 10^5$ W/cm$^2$. The value of $I_0$ is the saturation intensity of the transition 1-2 calculated at $B = 1$ T and assuming that $1/T = \gamma = 3\times 10^{13}$ s$^{-1}$. $I_1$ is set to satisfy $y = 0.6 x$. (b) Enlargement of (a) near the origin, which shows the intensity of the 4-wave mixing signal for a weak pump field. }
\end{figure}
This trend is reversed for a small incident pump intensity, when $\chi^{(3)} \sim 1/B$ and f(x, y) is nearly a constant. As a result, for weak pumps $I_3$ is higher in a smaller magnetic field, as illustrated in Fig.~4(b), which is the enlargement of Fig.~4(a) near the origin.  

\subsection{Stimulated Raman Scattering}

The very last term on the right-hand side of Eq.~(\ref{s43}) that we previously omitted describes another interesting nonlinear process: stimulated Raman Stokes scattering of the pump field $E_1(\omega_1)$ into the field $E_3(\omega_3)$; see Fig.~5. Note that this term does not depend on the second pump field $E_2(\omega_2)$; therefore for this section we can put $E_2 = 0$. In this case the amplitude of the off-diagonal density matrix element 
$\sigma_{43}$, which determines the optical polarization at the frequency of the nonlinear signal, becomes  
\begin{equation}
\sigma_{43} = \left(\frac{\Omega_{43}}{i\Gamma_{43}}n_{43}+\frac{|\Omega_{41}|^2\Omega_{43}}
{i^3\Gamma_{43}\Gamma^*_{31}\Gamma^*_{41}} n_{41}\right) / \left(1 + |\Omega_{41}|^2/(\Gamma_{43}\Gamma^*_{31})\right).
\end{equation}
Here the complex detuning at the difference frequency is given by  $\Gamma_{31} = \gamma_{31} + i(\omega_{31} -\omega_{1}+ \omega_3)$, whereas other detunings are still  $\Gamma_{41} = \gamma_{41} + i(\omega_{41} -\omega_{1})$ and $\Gamma_{43} = \gamma_{43} + i(\omega_{43} -\omega_3)$.
\begin{figure}[htb]
\centerline{
\includegraphics[width=7cm]{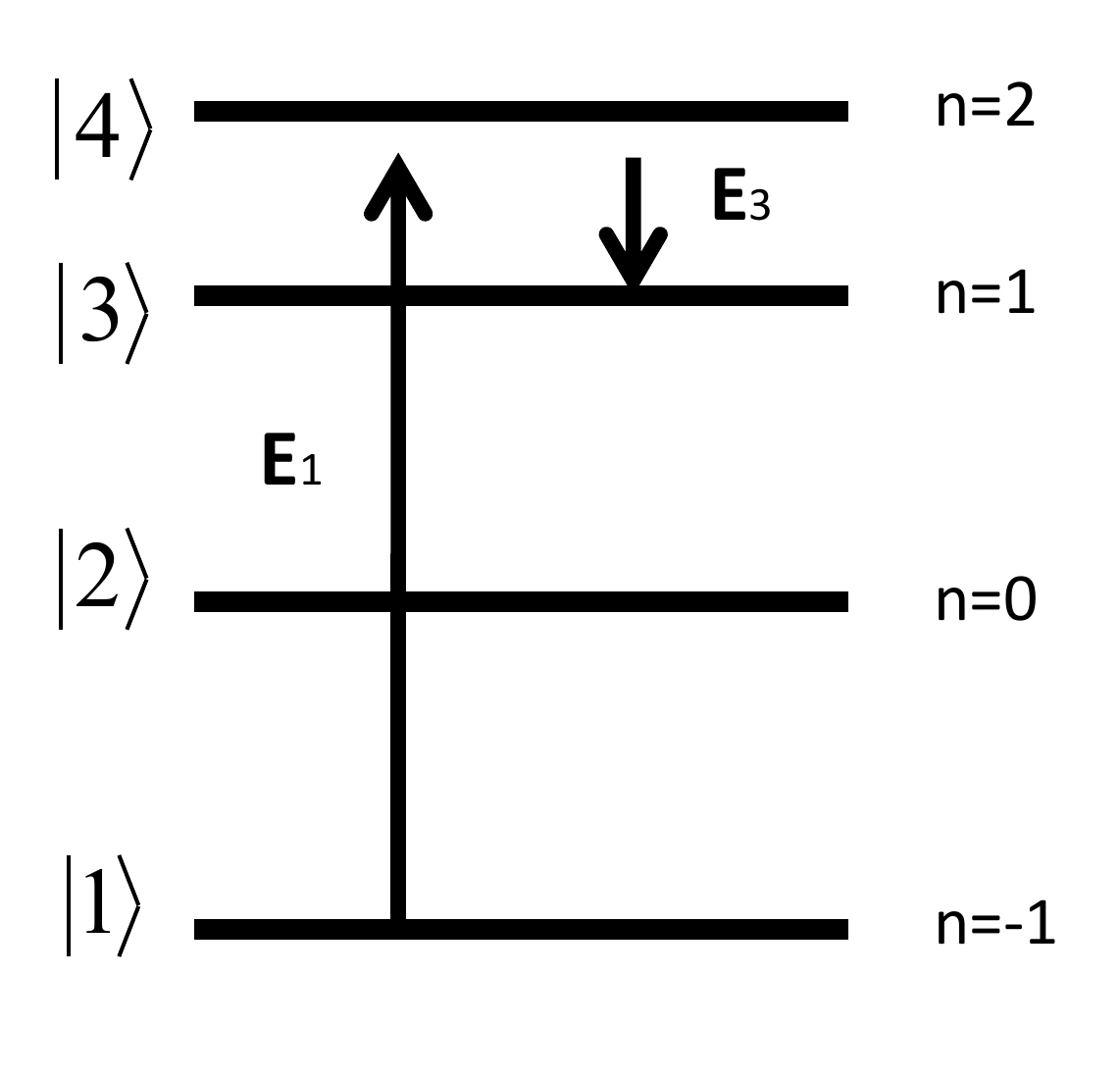}}
 \caption{Raman Stokes scattering of the incident field $E_{1}$ into the signal $E_3$.}
\end{figure}

Since the polarization $P(\omega_3)$ is proportional to the field $E_3$, the small-signal solution to the wave equation Eq.~(\ref{maxw}) has an exponential form, 
 $E_{3} = E_0 \exp{(g z)}$, where $g$ is given by 
\begin{equation}
g = \frac{2\pi\omega_3 N_{3D}\mu_{43}\tilde{\mu}_{43}}{\hbar c\Gamma_{43}} \left(n_{43}-\frac{|\Omega_{41}|^2}{\Gamma^*_{31}\Gamma^*_{41}} n_{41}\right) / \left(1 + |\Omega_{41}|^2/(\Gamma_{43}\Gamma^*_{31})\right), 
\label{gain1}
\end{equation}
and $N_{3D} = 2/(\Delta z\pi l_c^2)$ is the volume density of electrons in a layer of thickness $\Delta z$. The real part of $g$ gives the spectrum of the Raman gain. It is similar to the one derived for resonant Raman lasers in atomic and quantum-well systems \cite{belyanin2005,belyanin2006,rost}. The gain peaks at the frequency of the two-photon resonance $\omega_{1} - \omega_3 = \omega_{31}$. Its peak value increases when the pump frequency is tuned closer to the one-photon resonance $\omega_1 =  \omega_{41}$.

To estimate the maximum gain, we assume exact resonance for the pump and Stokes fields with corresponding transition frequencies $\omega_{41}$ and $\omega_{43}$, and take all dephasing rates to be the same, so that $\Gamma_{ij} \sim \gamma$. Then the gain factor is simplified to
\begin{equation}
g \Delta z \sim  \frac{4 \omega_3 \mu^2_{43}}{\hbar \gamma c l^2_c } \left(n_{43}+\frac{|\Omega_{41}|^2}{\gamma^2} n_{14}\right) / \left(1 + |\Omega_{41}|^2/\gamma^2\right).
\label{gain2}
\end{equation}
For a weak pumping $|\Omega_{41}|^2 \ll \gamma^2$, all population stays in the ground state of the system,
$n_{14} \sim 1$ and $n_{43} \sim 0$. Then the maximum gain becomes 
\begin{equation}
g \Delta z \sim  \frac{4 \omega_3 \mu^2_{43}}{\hbar \gamma c l^2_c } \frac{|\Omega_{41}|^2}{\gamma^2}. 
\label{small} 
\end{equation}

When expressed in usual dimensions cm/W for comparison with other materials,  the Raman gain coefficient is really huge: around 20 cm/W in the magnetic field of 1 Tesla, and assuming $\gamma = 3\times 10^{13}$ s$^{-1}$. This is many orders of magnitude higher gain than the one reported for resonant intersubband Raman scattering in conventional 2D semiconductor systems: asymmetric coupled quantum well systems or quantum cascade lasers \cite{belyanin2005,belyanin2006,liu,faist}. 
\begin{figure}[htb]
\centerline{
\includegraphics[width=15cm]{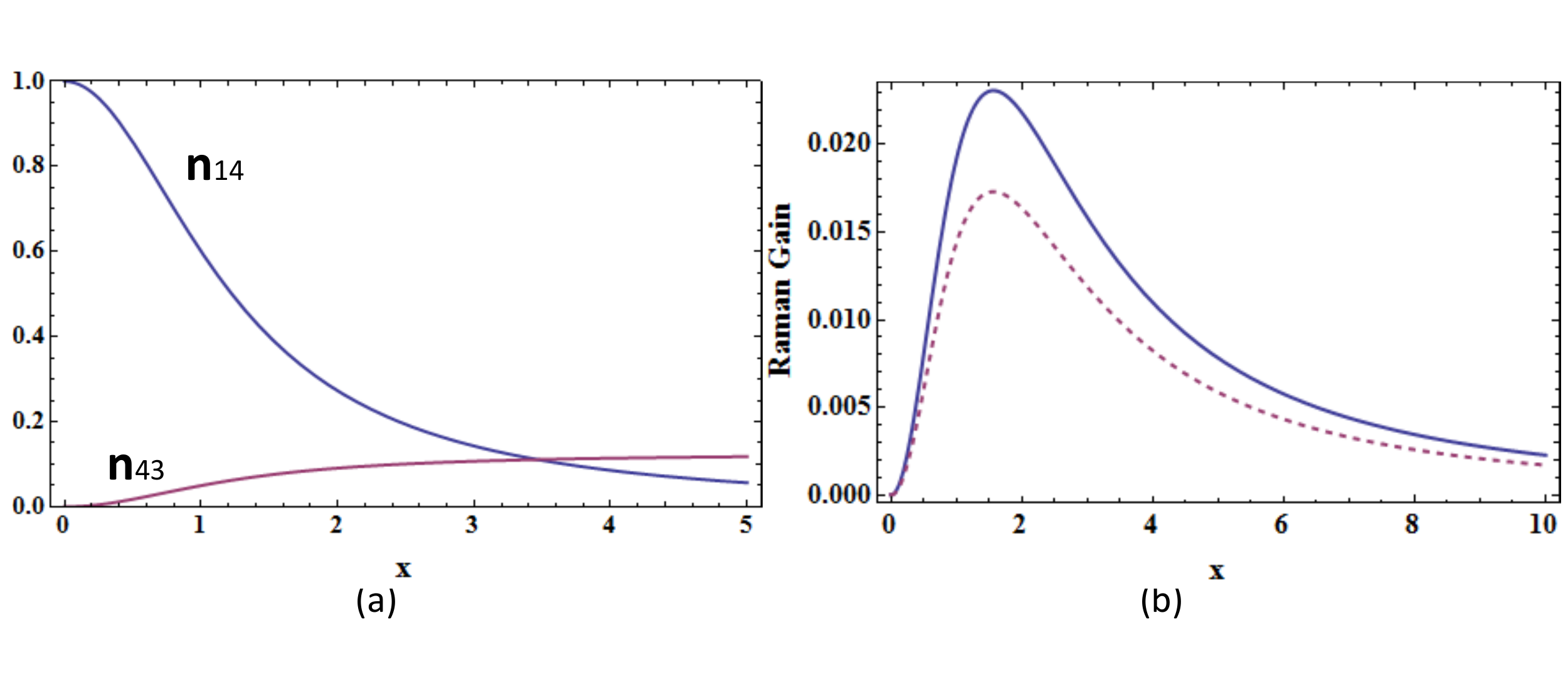}}
 \caption{(a) Population differences as functions of the normalized pump field. (b)  Gain $g \Delta z$ for one monolayer of graphene as a function of the normalized pump field in the magnetic field $B = 1$ T. Solid line: the total gain, dashed line: only the Raman part assuming $n_{43} = 0$. }
\end{figure}

 For a stronger pump field, effects of the optical pumping and saturation become important. From the structure of the gain expression Eq.~(\ref{gain2}), it is clear that the gain reaches a maximum value when the pump field is of the order of the saturation value. This is a generic property of all resonant nonlinearities. For even higher fields, the gain drops due to a decrease in $n_{14}$ and an additional power-broadening term $1+|\Omega_{41}|^2/\gamma^2$ in the denominator.   Using the same notation as in the  previous section, we define the saturation field $E^s_{41} = \hbar \sqrt{\gamma_{41}/T_{41}} /\mu_{41}$ and the dimensionless pump field $x = E_{1} / E^s_{41}$. Taking all relaxation times to be the same, $T_{ij} \sim T$, all population differences and the gain factor can be calculated analytically. They are shown in Fig.~6 as functions of the normalized pump field. Note that for our choice of equal relaxation rates, the optical pumping to the upper state 4 results in the population inversion on the signal transition: $n_{43} > 0$. This leads to an additional contribution to the gain, as is clear from comparing the total gain and the Raman contribution. The peak gain of about 2 \% is amazingly high for just one monolayer of the material. By stacking several layers and placing the system in a high-Q THz laser cavity one can achieve a THz Raman laser with emission wavelength tunable by a magnetic field.   
 
 In conclusion, we presented detailed studies of the linear and nonlinear optical response of graphene placed in a strong magnetic field. We showed that this system has an extremely high optical nonlinearity. We discussed two schemes of the nonlinear THz generation in graphene based on the resonant four-wave mixing and Raman scattering of intense mid-infrared fields. The predicted nonlinear power makes graphene interesting for a variety of THz applications. Furthermore, one expects to find a similar physics of the nonlinear optical interactions in topological insulators, where the surface states have a massless dispersion and demonstrate a similar magneto-absorption on the transitions between the Landau levels  \cite{topins1}. 

This work has been supported in part by NSF Grants OISE-0968405 and EEC-0540832, and by the NHARP Project No. 003658-0010-2009.

\end{document}